\documentclass{article}
\usepackage{IEEEtrantools}
\usepackage{eso-pic}

\AddToShipoutPictureBG*{%
  \AtPageLowerLeft{%
    \put(\LenToUnit{\paperwidth}+450,-610){%
      \makebox[-\textwidth][r]{%
        \raisebox{\dimexpr\textheight+\footskip-2\baselineskip}[0pt][0pt]{%
         To appear in \textit{Proc. ICASSP 2024, April 14-19, 2024, Seoul, Korea}     \hspace{23.5em}        \copyright~IEEE~\the\year
        }%
      }%
    }%
  }%
}
\usepackage{adjustbox}
\usepackage{cite}
\usepackage{booktabs}
\usepackage{spconf,amsmath,graphicx}
\usepackage{ amssymb }
\usepackage{array}
\usepackage{makecell}
\usepackage{fontawesome}
\usepackage{multirow}
\usepackage{footnote}
\makesavenoteenv{tabular}
\makesavenoteenv{table}

\ninept
\title{Parameter Efficient Finetuning for Speech Emotion Recognition and Domain Adaptation}
\name{Nineli Lashkarashvili, Wen Wu, Guangzhi Sun, Philip C. Woodland\thanks{Nineli Lashkarashvili is funded by DeepMind and Cambridge Trust. This work has been performed using resources provided by the Cambridge Tier-2 system operated by the University of Cambridge Research Computing Service (www.hpc.cam.ac.uk) funded by EPSRC Tier-2 capital grant EP/T022159/1. }}
\address{Cambridge University Engineering Dept., Trumpington St., Cambridge, CB2 1PZ U.K.\\
\small{\{nl438, ww368, gs534, pcw\}@eng.cam.ac.uk}}

\date{}
\begin{document}
%
\maketitle
\begin{abstract}
Foundation models have shown superior performance for speech emotion recognition (SER). 
However, given the limited data in emotion corpora, finetuning all parameters of large pre-trained models for SER can be both resource-intensive and susceptible to overfitting. 
This paper investigates parameter-efficient finetuning (PEFT) for  SER. 
Various PEFT adaptors are systematically studied for both classification of discrete emotion categories and prediction of dimensional emotional attributes. 
The results demonstrate that the combination of  PEFT methods surpasses full finetuning with a significant reduction in the number of trainable parameters. 
Furthermore, a two-stage adaptation strategy is proposed to adapt models trained on acted emotion data, which is more readily available, to make the model more adept at capturing natural emotional expressions. 
Both intra- and cross-corpus experiments validate the efficacy of the proposed approach in enhancing the performance on both the source and target domains.



\end{abstract}

\section{Introduction}
\label{sec:intro}

Emotion is a complex process influencing human interaction and cognitive processes~\cite{van2016social,tyng2017influences}. Speech emotion recognition (SER) enables systems to perceive and respond to emotional cues in spoken language and plays a crucial role in human-computer interaction. 
SER has garnered growing interest in recent years~\cite{Kim_2013, Poria2017,wu2021emotion}, yet the data scarcity remains a critical challenge for its advancement.
Two key concerns regarding emotion datasets are their limited corpus size and lack of naturalness.
Most emotional databases only contain a few hours of recordings~\cite{busso2008iemocap, livingstone2018ryerson, martin2006enterface, cao2014crema, burkhardt2005database} with many acquiring emotional speech by participants reading sentences in designated emotional states~\cite{zhou2022emotional, busso2008iemocap, livingstone2018ryerson, burkhardt2005database}. 
These acted portrayals tend to reflect more standardised behaviours rather than the nuanced and ambiguous emotional expressions encountered in everyday interactions.


The development of foundation models has paved the way to a new paradigm, enabling the customisation of general-purpose models for specific downstream tasks.
Foundation models are large machine learning models trained on a very large quantity of data usually by self-supervised learning. 
Transferring knowledge from pre-trained foundation models can help alleviate the data sparsity issue of downstream tasks~\cite{zhang2022bigssl,morais2022speech,wu2023self}. 
However, finetuning the whole foundation model is resource-consuming and can lead to overfitting and catastrophic forgetting. 
Parameter-efficient fine-tuning (PEFT) algorithms have been proposed to help mitigate these issues. 
PEFT approaches only fine-tune a small number of (extra/internal) model parameters while keeping most pre-trained foundation model parameters frozen. 
PEFT has been shown to be successful for natural language processing tasks~\cite{hu2021lora, houlsby2019parameter, zaken2021bitfit} and also some speech tasks~\cite{chen2023exploring, li2023evaluating}. 
However, the efficacy of PEFT methods for SER, including their effectiveness across discrete emotion classes and dimensional emotion attributes hasn't been extensively explored.

This paper provides a comprehensive exploration of PEFT methods for SER. 
Emotion states can be defined either by discrete emotion classes (i.e., happy, sad, angry, neutral)~\cite{plutchik1980general} or dimensional attributes (i.e., valence, arousal, dominance)~\cite{mehrabian1980basic}. 
Various PEFT methods, along with combinations thereof, are first employed to investigate both emotion classification and emotion attribute prediction, which aims to ascertain whether the methods exhibit consistent patterns of performance across both tasks. 

Furthermore, this paper investigates emotion domain adaptation via PEFT. 
SER datasets containing acted emotion data are more prevalent since acted emotion data are relatively easier to collect than natural emotion data. 
Therefore, the paper examines whether a model trained on acted emotion data can generalise to natural emotion. 
A two-stage adaption framework is developed for the adaptation of models trained on acted emotion data to the domain of natural emotions using the PEFT approach. 
Both intra- and cross-corpus adaptation are studied and results suggest that prediction of emotion in natural settings benefits from such adaptation.

The rest of the paper is organised as follows.
Section~\ref{sec: methods} introduces PEFT adaptors and the proposed approach. 
The experimental setup and results are presented in Section~\ref{sec: setup} and Section~\ref{sec: results} respectively, followed by conclusions.

\section{Methodology}
\label{sec: methods}
\subsection{Model structure}

The backbone structure follows an upstream-downstream framework~\cite{mohamed2022self}. 
Both wav2vec 2.0~\cite{baevski2020wav2vec}\footnote{https://huggingface.co/facebook/wav2vec2-base} and HuBERT~\cite{hsu2021hubert}\footnote{facebook/hubert-large-ll60k} are used as upstream architectures. 
Both models consist of a CNN feature extractor and Transformer encoder parts, with wav2vec 2.0 containing 12 transformer blocks with a hidden dimension of 768 and 8 attention heads, and HuBERT comprising 24 transformer blocks with a hidden dimension of 1024 and 16 attention heads.
The downstream model is constructed by two fully connected layers following the SUPERB benchmark setup~\cite{yang2021superb}.
For emotion attribute prediction valence, arousal and dominance are predicted together.

\begin{figure}[tb]
    \centering
    \includegraphics[width=.9\linewidth]{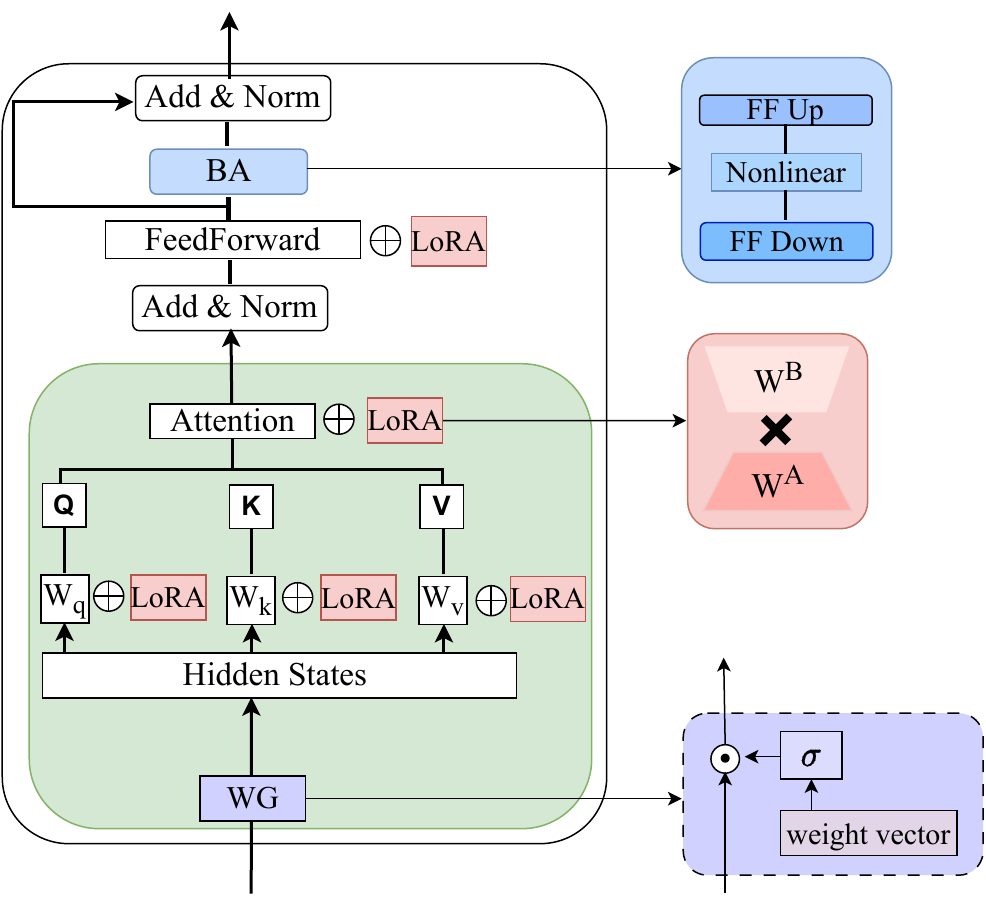}
    \caption{A wav2vec 2.0~/~HuBERT Transformer layer with various PEFT adaptors}
    \label{fig:model}
\end{figure}

\subsection{PEFT adaptors}
We first investigate the use of different PEFT adaptors and their combinations for SER. 
In the upstream model, PEFT modules are inserted in the Transformer encoders as shown in Fig.~\ref{fig:model}. The following adaptors are examined.

The Bottleneck Adaptor (BA)~\cite{houlsby2019parameter} is applied after the feed-forward layer of the Transformer encoder~\cite{yang2021superb}. 
It consists of a down-scaling layer which transforms a $d$-dimensional feature into $m$-dimensional $(m \ll d)$ (FF Down), a non-linear operation and an up-scaling layer which projects the $m$-dimensional feature back to $d$-dimensional (FF Up). 
A residual connection is added around the BA block. 
 
Low-Rank Adaptation (LoRA)~\cite{hu2021lora} adds trainable low-rank decomposition matrices to the weight matrices in the multi-head attention.
\begin{equation}
W_q' = W_q + \Delta W = W_q + W_q^BW_q^A  
\end{equation}
where $W_q \in \mathbb{R} ^{d\times k}$ is the original weight matrix for the query which is frozen during fine-tuning, $W_q^B\in \mathbb{R} ^{d\times r}$ and $W_q^A \in \mathbb{R}^{r \times k}$, $r \ll min(d,k)$ are the decomposition matrices which are updated during fine-tuning. 
Similar decompositions apply to weight matrices of the keys and values.

The Weighted Sum (WS)~\cite{yang2021superb} adaptor assigns learnable weights to the outputs of each Transformer block and returns the weighted sum for downstream modules. 
The scalar weights are passed to a softmax function to weight the importance of hidden states, which can also be treated as a PEFT method~\cite{chen2023exploring}.

In addition, this work proposes the Weight-Gating (WG) adaptor which assigns a learnable gating vector $\mathbf{g} \in \mathbb{R} ^d$ to each hidden state of dimension $d$ which is passed to a sigmoid activation function.
\begin{equation}
    h' = \sigma(\mathbf{g})\odot h
\end{equation}
where $h$ denotes the hidden states, $\sigma(\cdot)$ stands for sigmoid fuction,  and $\odot$ is element-wise product.

\subsection{Domain adaptation from acted emotion to natural emotion}
Acted (simulated) emotion data is relatively easier to collect compared to natural emotion and acted speech databases are thus more prevalent than natural emotion databases. 
Utterances in acted speech databases usually contain professional or semi-professional actor recordings who are instructed to express designated emotions. 
Sometimes, the participant reads the same scripted sentences in different emotions~\cite{zhou2022emotional, cao2014crema, livingstone2018ryerson, burkhardt2005database} 
\begin{figure}[tb]
    \centering
    \includegraphics[width=\linewidth]{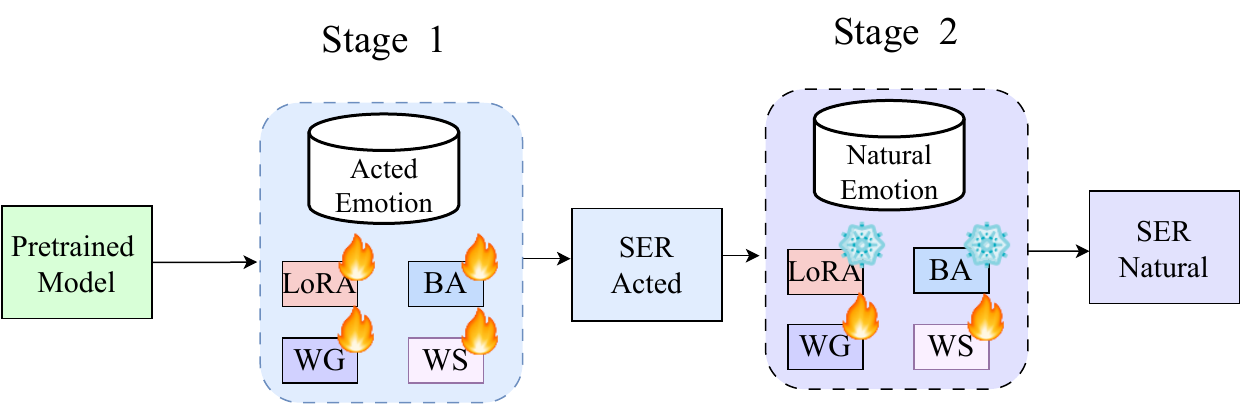}
    \caption{Proposed two-stage domain adaptation from acted to natural emotion prediction.
            \includegraphics[height=1.2em]{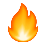} denotes that PEFT module is updated while \includegraphics[height=1.2em]{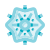} that it is frozen.
            The figure depicts the scenario when BA and LoRA weights from stage 1 are not updated.}
    \label{fig: adapt}
\end{figure}
which may not convey real-life emotions adequately, and can be exaggerated. Models trained on acted emotion data may not generalise well to real-world applications. 
Therefore, we investigate adapting models trained on acted emotion data to natural emotion data using PEFT. 
As shown in Fig.~\ref{fig: adapt}, the pipeline contains two stages. 
In the first stage, the pre-trained upstream model is fine-tuned on acted emotion data using a combination of PEFT adaptors, which aims to leverage generic speech knowledge to SER. 
In the second stage, the SER model trained on acted emotion data 
is further fine-tuned on natural emotion data. 
Part of the PEFT adaptors are also frozen in the second stage and whether the adaptation can improve the performance in the target domain while avoiding catastrophic forgetting in the source domain is of interest.
The WG and WS adaptors are always updated during training, reflecting their role in weighting the significance of hidden states, and they should align with the specific task at hand.

\begin{table*}[!h]
    \centering
    
     \begin{adjustbox}{max width=\linewidth, scale=1}
    
    \begin{tabular}{cccc|ccccc|ccccc}
    \toprule
    &&& & \multicolumn{5}{c|}{$\mathrm{wav2vec\ 2.0_{BASE}}$}  &  \multicolumn{5}{c}{$\mathrm{HuBERT_{LARGE}}$}\\
 BA & LoRA  &WS& WG  & \# param & Acc(\%) & $\mathrm{CCC_V}$ & $\mathrm{CCC_A}$ & $\mathrm{CCC_D}$ & \# param & Acc(\%)  & $\mathrm{CCC_V}$ & $\mathrm{CCC_A}$ & $\mathrm{CCC_D}$\\
 \midrule
    $\checkmark$  & & &  & 1.2 M & 65.34 & 0.630 & 0.705 & 0.514 & 3.2 M & 69.83 & 0.664 & 0.714 & 0.524 \\
    
    & $\checkmark$ & & & 1.3 M & 64.89 &0.599 & 0.693  & \textbf{0.523} & 3.5 M& 65.16 & 0.579 & 0.702 & \textbf{0.560} \\
      & &$\checkmark$ & & 12 & 59.13 & 0.428 & 0.673 & 0.485 & 24 & 65.32 & 0.514 & 0.692 & 0.552 \\
      & & & $\checkmark$& 9 K  & 52.58 & 0.375 & 0.646 & 0.495  & 25 K &  60.07 & 0.351 & 0.441 & 0.382\\
    \midrule
    $\checkmark$& $\checkmark$  & &      &  2.5 M    & 64.04 & \textbf{0.644} & 0.699 & 0.508 & 6.7 M & 64.56 & \textbf{0.678} & \textbf{0.724} & 0.542\\
    $\checkmark$& $\checkmark$  &  $\checkmark$  &       &  2.5 M  & 65.96 &  0.596 & \textbf{0.709} &  0.518& 6.7 M &  71.50 & 0.665 & 0.681 & 0.520   \\
    $\checkmark$& $\checkmark$   & $\checkmark$&   $\checkmark$    &  2.5 M   & \textbf{67.00} & 0.571 & {0.708} & 0.516 & 6.7 M & \textbf{71.88} & 0.651 & 0.701 & 0.521  \\
    \midrule
    \multicolumn{4}{c|}{FT} & 90 M & 66.30 &0.639 & 0.702 & 0.505 & 311 M &  68.53 & 0.652 & 0.715 & 0.527  \\
    \multicolumn{4}{c|}{PT} & 0 & 51.02 & 0.356 & 0.635 & 0.495 & 0 & 60.09 & 0.458 & 0.554 & 0.477 \\

    \bottomrule
\end{tabular}
    \end{adjustbox}

    \caption{IEMOCAP 5-cv mean accuracy and CCC scores for categorical and dimensional emotion prediction. `$\checkmark$' indicates the use of the PEFT adaptor during training. \# param denotes trainable parameters excluding downstream layers and BA, LoRA, WS, WG represent Bottleneck, Low-Rank Adaptation, Weighted Sum, and Weight-Gating adaptors, respectively.}
    \label{tab:peft_ser}

\end{table*}

\section{Experimental setup}
\label{sec: setup}

\subsection{Datasets}

IEMOCAP~\cite{busso2008iemocap} consists of 5 sessions. 
Each session contains scripted (acted) and improvised dialogues between two speakers. 
In the improvised part actors were asked to think of any past memory and express themselves for a certain emotion.
The IEMOCAP creators believe that such a setup will convey emotions closer to real life.
Both categorical labels and dimensional (valence-arousal-dominance) are provided. 
Following prior work~\cite{yang2021superb}, four emotion classes for categorical emotion prediction were used: happy (merged with excited), sad, angry, and neutral, resulting in 5,531 utterances. 
For dimensional emotion descriptor prediction, the entire dataset comprising 10,039 data samples were used.
A leave-one-session-out 5-fold cross-validation (5-cv) setup was used and the average scores are reported for classification and regression tasks.

\subsection{Implementation details}
The system is implemented in PyTorch. Models were trained for $20$ epochs using Adam optimiser with an initial learning rate of $5\times 10^{-4}$ for all experiments except finetuning ($5 \times 10^{-5}$) and a linear learning rate scheduler.
The best models were saved based on validation set performance and then used to evaluate on the test set. 
Following~\cite{yang2021superb}, discrete emotion classification was evaluated by accuracy (Acc) and emotion attributes prediction by the concordance correlation coefficient (CCC):
\begin{equation}
\label{eq:ccc}
    \mathrm{CCC} = \frac{2\rho\sigma_a\sigma_p}{\sigma_a^2 + \sigma_p^2 + (\mu_a - \mu_p)^2}
\end{equation}
where $\rho$ is the Pearson correlation coefficient between true label (t) and model prediction (p). $\sigma_a^2$, $\mu_a$ and $\sigma_p^2$, $\mu_p$ are the variances and means for annotation and predictions.


    
    

\section{Results}
\label{sec: results}

PEFT adaptors were compared to the pre-trained foundation model without any finetuning (PT) and fully fine-tuning (FT) where the CNN feature extractor is frozen and all transformer blocks are updated.

\subsection{Efficient finetuning for SER}
\begin{table*}[!h]
    \centering
    \ninept
      \begin{adjustbox}{max width=\linewidth, scale=1}
    
    \begin{tabular}{ccc|cccc|cc|cc}
\toprule
      
      &&& &&&& \multicolumn{2}{c|}{$\mathrm{wav2vec\ 2.0_{BASE}}$}  &  \multicolumn{2}{c}{$\mathrm{HuBERT_{LARGE}}$}\\
       & Source & Target &BA & LoRA  &WS& WG  &  $\mathrm{IEM_{sc}}$ & $\mathrm{IEM_{im}}$   &  $\mathrm{IEM_{sc}}$ & $\mathrm{IEM_{im}}$ \\
       \midrule
       \multirow{2}{3em}{Stage 1}&PT & IEM$_\text{sc}$ &$\checkmark$ &$\checkmark$&$\checkmark$&$\checkmark$  &      57.98  &  \underline{48.37} &  \textbf{71.63} & \underline{58.34}   \\
        &PT & IEM$_\text{im}$& $\checkmark$ &$\checkmark$&$\checkmark$&$\checkmark$&       \underline{52.13}   & 67.56 &  \underline{58.51} & 68.71  \\
         \midrule
         \multirow{4}{3em}{Stage 2}&IEM$_\text{sc}$ &  IEM$_\text{im}$ &$\ast$&$\ast$&$\checkmark$&$\checkmark$&  53.37 & 64.72 & 66.31 & 74.47 \\
         &IEM$_\text{sc}$ &  IEM$_\text{im}$ &  $\checkmark$&$\ast$&$\checkmark$&$\checkmark$&  55.32 & 68.91 & 63.83 & 74.66  \\
          &IEM$_\text{sc}$ & IEM$_\text{im}$ & $\ast$ &$\checkmark$&$\checkmark$&$\checkmark$&   \textbf{58.87} & 69.48 & 63.12 & 73.32\\
          &IEM$_\text{sc}$ & IEM$_\text{im}$ & $\checkmark$ &$\checkmark$&$\checkmark$&$\checkmark$&   57.98 & \textbf{70.63} & 61.88 & \textbf{74.66}\\

    \bottomrule
    \end{tabular}
    \end{adjustbox}
    \caption{
     Intra-corpus IEMOCAP scripted ($\mathrm{IEM_{sc}}$) to improvised ($\mathrm{IEM_{im}}$) adaptation.
     `$\checkmark$' denotes that PEFT module is updated and `$\ast$' that it is frozen.
     zero-shot results are underlined.
    }
    \label{tab:iem_sc_imp}
\end{table*}
Table~\ref{tab:peft_ser} presents the outcome of applying LoRA, BA, WS, and WG PEFT methods and their combinations to both categorical and dimensional emotion prediction tasks for the wav2vec 2.0 and HuBERT models. 
Notably, all PEFT methods, except for WG in HuBERT, exhibit superior performance compared to the pre-trained baseline. 
Augmenting the combination of BA, LoRA, and WS with WG yields an overall performance boost that surpasses fine-tuning (FT) while using 2-3\% of the total trainable parameters for both models. 
In terms of classification tasks, wav2vec 2.0 and HuBERT yield similar performance for different PEFT methods. 
HuBERT stands out particularly in categorical label prediction, achieving results not far from the state-of-the-art of 74.2\% \cite{gat2022speaker} without relying on external features, using only raw audio data.

However, a noteworthy observation is that the combination of LoRA and BA falls short of the best-performing standalone methods. One plausible explanation could be over-parameterisation, which may impede model generalisation and increase complexity, potentially leading to overfitting the training data. WS and WG seem to serve as corrective measures, potentially compensating for lost discriminative information and effectively modulating crucial features. Consequently, they exhibit the highest accuracies for each model.

The outcomes for dimensional emotional attribute prediction deviate from those of categorical classification. 
This difference can be attributed to the inherent dissimilarities between these tasks. 
Estimating continuous emotional attributes is intrinsically more challenging, and unlike categorical label prediction, the combination of BA and LoRA does not lead to overparameterisation.
Instead, this combination proves sufficient to capture the intricate relationships between the acoustic features and continuous emotional dimensions. 
For wav2vec 2.0 the addition of WS and WG was advantageous for arousal scores.
This could be attributed to the contribution of WS and WG which both appear to enhance the performance compared to PT for wav2vec 2.0.
Conversely, for HuBERT, this phenomenon doesn't hold and the performance declines when WS and WG are introduced.
The decrease in performance could be due to overmodulation, where the added complexity of these PEFT methods result in reduced performance.
It's worth noting that LoRA proves effective for dominance prediction, while BA excels in valence and arousal prediction. 
The results suggest that due to the increased task difficulty, PEFT modules tend to prioritize the enhancement of a specific emotion dimension. 
Overall, the combination of PEFT methods demonstrated superior performance compared to FT.
Furthermore, PEFT for both wav2vec 2.0 and HuBERT models for dimensional emotional attribute prediction surpassed existing state-of-the-art CCC scores for valence (0.582), arousal (0.667) and dominance (0.545)  \cite{srinivasan2022representation} with significantly fewer trainable parameters.


\subsection{Adaptation from acted emotion to natural emotion}
This section investigates the generalisability of a model trained on acted emotion data to natural emotion. 
The combination of BA, LoRA, WS and WG was used for stage 1 adapting since it produces the best performance in Table~\ref{tab:peft_ser}. 
Both intra- and cross-corpus adaptation were investigated. Fold 1 (the model trained on sessions 2-5 and tested on session 1) was used for IEMOCAP.


\begin{table*}[!h]
    \centering
      \begin{adjustbox}{max width=\linewidth}
    
    \begin{tabular}{ccc|cccc|cc|cc}
\toprule
      
      &&& &&&& \multicolumn{2}{c|}{$\mathrm{wav2vec\ 2.0_{BASE}}$}  &  \multicolumn{2}{c}{$\mathrm{HuBERT_{LARGE}}$}\\
       &Source & Target &BA & LoRA  &WS& WG  &  $\mathrm{ESD}$ & $\mathrm{IEM_{im}}$   &  $\mathrm{ESD}$ & $\mathrm{IEM_{im}}$ \\
       \midrule
        \multirow{2}{3em}{Stage 1}&PT & ESD &$\checkmark$ &$\checkmark$&$\checkmark$&$\checkmark$  & \textbf{91.58} & \underline{33.01} & \textbf{96.08} & \underline{50.67} \\
        &PT & $\mathrm{IEM_{im}}$& $\checkmark$ &$\checkmark$&$\checkmark$&$\checkmark$& \underline{36.33} & 67.56 & \underline{35.42} & 68.71 \\
        \midrule 
         \multirow{4}{3em}{Stage 2}&ESD & $\mathrm{IEM_{im}}$ & $\ast$&$\ast$&$\checkmark$&$\checkmark$&72.41 & 64.10 & 48.25 & 68.52 \\
         &ESD & $\mathrm{IEM_{im}}$ & $\checkmark$&$\ast$&$\checkmark$&$\checkmark$& 54.75 & \textbf{70.63} & 36.16 & \textbf{74.66}\\
        &ESD & $\mathrm{IEM_{im}}$ & $\ast$ &$\checkmark$&$\checkmark$&$\checkmark$& 62.25& 67.56 & 44.75 & 70.83  \\
        &ESD & $\mathrm{IEM_{im}}$ & $\checkmark$ &$\checkmark$&$\checkmark$&$\checkmark$&  56.08 & 69.28 & 35.83& 74.47 \\
    \bottomrule
    \end{tabular}
    \end{adjustbox}
    \caption{Cross-corpus ESD to IEMOCAP improvised ($\mathrm{IEM_{im}}$) adaptation. `$\checkmark$' denotes that PEFT module is updated and `$\ast$' that it is frozen.
     zero-shot results are underlined.
    }
    \label{tab: esd im}
\end{table*}
\subsubsection{Intra-corpus adaptation}
IEMOCAP contains scripted dialogues ($\mathrm{IEM_{sc}}$) with acted emotion and improvised ($\mathrm{IEM_{im}}$) which contain more natural emotional expression. 
We first investigated adapting a model trained on scripted data to improvised emotion.

The first two rows of Table~\ref{tab:iem_sc_imp} show the results of directly fine-tuning pre-trained model on scripted data and improvised data respectively. 
It can be seen that achieving high accuracy on scripted emotional state prediction is challenging for the wav2vec 2.0 model, while HuBERT exhibits better generalisation and decent performance across different data segments. 
HuBERT representations are more powerful which was also evident in PEFT SER.
The last four rows present the results of stage 2 adaptation. 
Overall, adapting a model trained on scripted data using improvised data enhances SER performance for natural emotions. 
There's a consistent improvement in target domain performance for both wav2vec 2.0 and HuBERT models.
The accuracy of improvised data surpasses the baseline achieved by direct fine-tuning on improvised data, indicating that the model benefits from leveraging knowledge from acted emotion. 
Adapting a HuBERT model trained on scripted data to the improvised dataset results in a performance boost of about 6\% on the improvised set.
Freezing a portion of PEFT modules has a minimal or even beneficial impact on the target task's performance, notably enhancing prediction quality on the source domain. 
This phenomenon is particularly pronounced in the case of HuBERT, where fine-tuning only WS and WG modules yields comparable performance to fine-tuning all PEFT modules and avoids forgetting of source task.

\subsubsection{Cross-corpus adaptation}
The use of an external ESD corpus was investigated for cross-corpus adaptation.
In ESD~\cite{zhou2022emotional} speakers were directed to read each sentence for all designated emotion categories.  
For ESD, English utterances corresponding to the four emotion classes in IEMOCAP and original train/evaluation/test splits were used.

Table~\ref{tab: esd im} presents the results of adapting from the ESD dataset to the IEMOCAP improvised set. 
Underlined values in the table highlight that a model trained on one corpus faces challenges generalising to the other. 
However, it's worth noting that HuBERT, which was only trained on the ESD data, exhibits reasonable accuracy on the improvised set. 
During the stage 2 adaptation process, when selected adaptors were frozen, a trade-off is observed between preserving knowledge from the source domain and effectively adapting to the target domain.
Both models exhibit the best improvised accuracy when LoRA is frozen and and their performance across different methods is quite similar.
Presumably, LoRA captures important features and freezing it allows valuable knowledge to be retained and doesn't hinder generalisation.
The results suggest that freezing part of the PEFT modules similar to intra-corpus adaptation is beneficial in preserving source task accuracy.
Unlike intra-corpus adaptation, a large decline in performance is observed for the ESD dataset after adaptation, possibly due to the fact that emotions in ESD follow a more paradigmatic pattern.

\section{Conclusion}
This work investigates PEFT for speech emotion recognition. 
Results show that PEFT can achieve comparable performance to full fine-tuning while requiring far fewer updated parameters. 
In both categorical and dimensional emotional descriptor prediction tasks, the combination of various PEFT adaptors results in enhanced system performance, demonstrating comparable or even superior outcomes to the state-of-the-art models, all while significantly reducing trainable parameters and without relying on external features beyond raw audio data.
Furthermore, a two-stage pipeline is introduced to adapt models trained on acted emotion to better handle natural emotions using PEFT. 
The results from both intra- and cross-corpus experiments highlight the effectiveness of this adaptation strategy, showcasing performance enhancements in both source and target domains.
Freezing BA or LoRA modules was found helpful in avoiding catastrophic forgetting of the source domain. 
This approach not only facilitates the retention of the source task but also offers the prospect of achieving performance levels equivalent to, if not surpassing, a model trained directly on natural emotion.

\bibliographystyle{IEEEbib}
\bibliography{main}

\end{document}